\documentclass[aps,showpacs]{revtex4}
\usepackage[dvips]{graphicx}
\usepackage{amsmath}

\begin{document}
\title{London's Equation from Abelian Projection}
\author{Vladimir Dzhunushaliev}
\email{dzhun@hotmail.kg}
\affiliation{Dept. Phys. and Microel. Engineer., Kyrgyz-Russian
Slavic University, Bishkek, Kievskaya Str. 44, 720000, Kyrgyz
Republic}
\author{Douglas Singleton}
\email{dougs@csufresno.edu}
\affiliation{Physics Dept., CSU Fresno, 2345 East San Ramon Ave.
M/S 37 Fresno, CA 93740-8031, USA}

\date{\today}

\begin{abstract}
Confinement in non-Abelian gauge theories, such as QCD,
is often explained using an analogy to type II
superconductivity. In this analogy the existence of the
``Meissner'' effect for quarks with respect to the QCD vacuum
is an important element. Here we show
that using the ideas of Abelian projection it is possible to arrive
at an effective London equation from a non-Abelian gauge theory.
(London's equation gave a phenomenological description of
the Meissner effect prior to the Ginzburg-Landau
or BCS theory of superconductors). The
Abelian projected gauge field acts as the E\&M field in normal
superconductivity, while the remaining non-Abelian components
form a gluon condensate which is described via an effective scalar
field. This effective scalar field plays a role similar
to the scalar field in Ginzburg-Landau theory.
\end{abstract}


\maketitle

\section{Introduction}

One of the difficult aspects of QCD is that it
is a nonlinear theory. At the classical level one writes down
the field equations of the nonlinear theory, and by inspiration
or numerically solves the field equations. For non-Abelian gauge
theories this leads to interesting classical field configurations
such as monopoles, merons, instantons {\it etc.}
In the quantized version of these theories there are
suggestions of interesting configurations like those
listed above. For example, the chromoelectric flux tubes
that are hypothesized to exist between quarks in the dual,
color superconductivity picture of confinement.

Quantizing nonlinear field theories presents additional
challenges and questions. In the case of general relativity
no complete quantum version of the theory exists. In the case
of QCD a quantum version of the theory exists, but
because the QCD coupling constant is large, the standard perturbative
quantum field theory techniques do not apply in the low energy
regime. This means that
getting numerical predictions out of QCD is difficult. At present
there is no universal analytical technique for handling detailed
questions about QCD (this is in contrast with QED and the theory
of the electroweak theory where the techniques of perturbative
quantum field theory supply a general method for calculating
results analytically). The nonperturbative aspects of QCD are dealt
with using numerical techniques \cite{chernodub}.

In the 1950's Heisenberg and co-workers proposed a scheme for
investigating quantum, nonlinear spinor fields \cite{heisenberg}, which
yielded interesting results concerning the spin
and charge of the electron. This nonperturbative quantization technique
distinguished between field operators of interacting fields and
field operators of noninteracting fields (Heisenberg's ideas
were developed before the widespread adoption of the path
integral techniques for quantizing field theories and as such
are given in terms of field operators). The algebra obeyed by the
noninteracting field operators is just the standard canonical
commutation relationships (see for example pg. 86 of ref.
\cite{mandl}). The algebra for the interacting fields, on the
other hand, is determined from the Green's functions for the
field operators of the nonlinear fields,
$\mathcal{G}(x_1, x_2 \ldots , x_n) = \langle Q |
\hat{A}(x_1), \hat{A}(x_2), \ldots \hat{A}(x_n) | Q \rangle$,
where $\hat{A}(x_i)$ is the field operator at position $x_i$;
$\mathcal{G}(x_1, x_2 \ldots , x_n)$ is the Green's function;
$|Q\rangle$ is a quantum state. Conversely if the
$\mathcal{G}(x_1, x_2 \ldots , x_n)$s are known then this
gives the quantum states $|Q\rangle$. The Green's functions
are determined from the field equations in the following way:
one starts with the classical field equations and turns the
classical fields into field operators. One uses these
operator field equations to determine the
lowest order Green's functions. Next one applies various order
derivative operators on these operator
field equations to obtain equations which have higher
order powers of the field operators, and which yield the
higher order Green's functions. In this way an infinite set of
coupled, differential equations are constructed
which connect all orders of Green's functions. In general
it is not possible to solve this system of differential
equations analytically, so approximation techniques must
be used.
\par
Here we apply a version of Heisenberg's method to non-Abelian
gauge theory. In our treatment we assume that the non-Abelian
gauge fields can be separated into two classes: stochastic,
disordered fields and ordered fields. This is similar
to Abelian Projection ideas \cite{hooft}
where the diagonal gauge fields associated with the Abelian subgroup
are ordered (an example of this ordering would be a non-zero
expectation, $\langle A^{diag}_z \rangle \neq 0$, inside a
flux tube stretched between quarks) and with the off-diagonal gauge
fields being disordered so that $\langle A^{off}_\mu \rangle = 0$
outside of the flux tube (but $\langle (A^{off}_\mu)^2 \rangle \neq 0$
outside the flux tube). This picture also has some common
points with the work of Nielsen, Olesen and others
\cite{no1} - \cite{amb} on the response of the QCD vacuum to
a homogeneous color magnetic field, $H$. In ref. \cite{no1} it was
shown that a homogeneous SU(2) magnetic field lowered the
energy of the vacuum, but was unstable. This instability was
removed by the formation of domains \cite{no2}. These
domains took the form of SU(2) magnetic flux tubes which formed
a random SU(2) magnetic quantum liquid \cite{amb} with the
property that $\langle H \rangle =0$ and
$\langle H^2 \rangle \neq 0$. The difference with our approach
outlined above is that we assume the expectation values are in
terms of the gauge potentials rather than the magnetic field --
$\langle A^{off}_\mu \rangle = 0$ and $\langle (A^{off}_\mu)^2 \rangle
\neq 0$. Also, in line with the ideas of Abelian Projection, the
diagonal, Abelian potentials ($A^{diag}_z$) and the off-diagonal,
non-Abelian potentials ($A_{\mu} ^{off}$) play different roles
in the present paper.
\par
In applying these ideas to an SU(2)
gauge theory we assume that the off-diagonal components form a
condensate which can be described by an effective scalar field
similar to that in Ginzburg-Landau theory \cite{gl}. This
is also related to work by Cornwall \cite{cw} \cite{cw1} where a
connection is established between a scalar field and the expectation
of the Yang-Mills field strength tensor squared -- $\phi (x)
\leftrightarrow \langle Tr G_{\mu \nu} ^2 \rangle$. In the
present work we follow the Abelian Projection ideas and
assign different roles to the Abelian and the off diagonal,
non-Abelian gauge potentials. Only the off-diagonal gauge
potentials are assumed be involved in the condensate. With
respect to this condensate the field equation
for the Abelian component then takes the form
of the London equation, so that the Abelian field develops
a mass as it penetrates into regions characterized by the
a non-zero value for off-diagonal
condensate. The mass of the Abelian field and the
assumed effective mass of the condensate are not the same, which
is in contrast to refs. \cite{cw} \cite{cw1} where all the SU(2)
gauge potentials play an equivalent role, leading to a single
common effective mass for all the gluons.

\section{Separation of Components}

In this section we follow the conventions of
Ref. \cite{kondo}. Starting with the SU(N)
gauge group with generators $T^B$ we define the SU(N)
gauge fields, $\mathcal{A}_\mu=
\mathcal{A}^B_\mu T^B$. Let $G$ be a subgroup
of SU(N) and $SU(N)/G$ is a coset. Then the gauge field
$\mathcal{A}_\mu$ can be decomposed as
\begin{eqnarray}
  \mathcal{A}_\mu & = & \mathcal{A}^B_\mu T^B = a^a_\mu T^a + A^m_\mu T^m ,
\label{sec2-10a}\\
  a^a_\mu & \in & G \quad \text{and} \quad A^m_\mu \in SU(N)/G
\label{sec2-10b}
\end{eqnarray}
where the indices $a,b,c \ldots $ belongs to the subgroup G and
$m,n, \ldots $to the coset SU(N)/G; $B$ are SU(N) indices.
Based on this the field strength can be decomposed as
\begin{equation}
  \mathcal{F}^B_{\mu\nu} T^B = \mathcal{F}^a_{\mu\nu}T^a +
  \mathcal{F}^m_{\mu\nu}T^m
\label{sec2-20}
\end{equation}
where
\begin{eqnarray}
  \mathcal{F}^a_{\mu\nu} & = & \phi ^a_{\mu\nu} + \Phi^a_{\mu\nu}
  \; \; \; \in G ,
\label{sec2-30a}\\
  \phi ^a_{\mu\nu} & = & \partial_\mu a^a_\nu - \partial_\nu a^a_\mu +
  f^{abc}a^b_\mu a^c_\nu \; \; \; \in G ,
\label{sec2-30b}\\
  \Phi^a_{\mu\nu} & = & f^{amn} A^m_\mu A^n_\nu \; \; \; \in G ,
\label{sec2-30c}\\
  \mathcal{F}^m_{\mu\nu} & = & F^m_{\mu\nu} + G^m_{\mu\nu} \; \; \;
  \in SU(N)/G ,
\label{sec2-30d}\\
  F^m_{\mu\nu} & = & \partial_\mu A^m_\nu - \partial_\nu A^m_\mu +
  f^{mnp} A^n_\mu A^p_\nu \; \; \; \in SU(N)/G ,
\label{sec2-30e}\\
  G^m_{\mu\nu} & = & f^{mnb}
  \left(
  A^n_\mu a^b_\nu - A^n_\nu a^b_\mu
  \right) \; \; \; \in SU(N)/G
\end{eqnarray}
where $f^{ABC}$ are the structural constants of
SU(N). The SU(N) Yang-Mills field equations can be decomposed
as
\begin{eqnarray}
  d_\nu \left( \phi ^{a\mu\nu} +\Phi^{a\mu\nu} \right) & = &
  -f^{amn} A^m_\nu
  \left(
  F^{n\mu\nu} + G^{n\mu\nu}
  \right) ,
\label{sec2-40a}\\
  D_\nu \left( F^{m\mu\nu} + G^{m\mu\nu} \right) & = &
  - f^{mnb}
  \left[
  A^n_\nu \left( \phi ^{b\mu\nu} + \Phi^{b\mu\nu} \right) -
  a^b_\nu \left( F^{n\mu\nu} + G^{n\mu\nu} \right)
  \right]
\label{sec2-40b}
\end{eqnarray}
where $d_\nu [\cdots]^a = \partial_\nu [\cdots]^a +
f^{abc} a^b_\nu [\cdots]^c$ is the covariant derivative on the
subgroup G and
$D_\nu [\cdots]^m = \partial_\nu [\cdots]^m +
f^{mnp} A^n_\nu [\cdots]^p$

Specializing to the SU(2) case we let $SU(N) \rightarrow SU(2)$ ,
$G \rightarrow U(1)$, and $f^{ABC} \rightarrow \epsilon ^{ABC}$.
Setting the indices as $a = 3$ and consequently $m,n = 1,2$,
our classical equations become
\begin{eqnarray}
  \partial_\nu \left( \phi ^{\mu\nu} +\Phi^{\mu\nu} \right) & = &
  -\epsilon^{3mn} A^m_\nu
  \left(
  F^{n\mu\nu} + G^{n\mu\nu}
  \right),
\label{sec3-10a}\\
  D_\nu \left( F^{m\mu\nu} +G^{m\mu\nu} \right) & = &
  -\epsilon^{3mn}
  \left[
  A^n_\nu \left( \phi ^{\mu\nu} + \Phi^{\mu\nu} \right) -
  a_\nu \left( F^{n\mu\nu} + G^{n\mu\nu} \right)
  \right]
\label{sec3-10b}
\end{eqnarray}
Since $G=U(1)$ we have $d_\nu = \partial_\nu$.

\section{Heisenberg Quantization}

In quantizing the classical system given in Eqs. (\ref{sec3-10a}) -
(\ref{sec3-10b}) via Heisenberg's method one first replaces the
classical fields by field operators $a_{\mu} \rightarrow \hat{a}_\mu$
and $A^m_{\mu} \rightarrow \hat{A}^m_\mu$. This yields the
following differential equations for the operators
\begin{eqnarray}
  \partial_\nu \left( \hat{\phi}^{\mu\nu} + \hat{\Phi}^{\mu\nu} \right) & = &
  -\epsilon^{3mn} \hat A^m_\nu
  \left(
  \hat F^{n\mu\nu} + \hat G^{n\mu\nu}
  \right),
\label{sec4-10a}\\
  D_\nu \left( \hat F^{m\mu\nu} + \hat G^{m\mu\nu} \right) & = &
  -\epsilon^{3mn}
  \left[
  \hat A^n_\nu \left( \hat \phi^{\mu\nu} + \hat \Phi^{\mu\nu} \right) -
  \hat a_\nu \left( \hat F^{n\mu\nu} + \hat G^{n\mu\nu} \right)
  \right]
\label{sec4-10b}
\end{eqnarray}
These nonlinear equations for the field operators of
the nonlinear quantum fields can be used to determine
expectation values for the field operators $\hat a_\mu$ and
$\hat A^m_\mu$ ({\it e.g.} $\langle \hat a_\mu \rangle$, where
$\langle \cdots \rangle = \langle Q | \cdots | Q \rangle$ and
$| Q \rangle$ is some quantum state). One can also use these
equations to determine the expectation values of operators
that are built up from the fundamental operators $\hat a_\mu$
and $\hat A^m_\mu$. For example, the ``electric'' field
operator, $\hat E_z = \partial _0 \hat a_z - \partial _z \hat a_0$
giving the expectation $\langle \hat E_z \rangle$.
The simple gauge field expectation values,
$\langle \mathcal{A}_\mu (x) \rangle$, are obtained by
average Eqs. \eqref{sec4-10a} \eqref{sec4-10b} over some
quantum state $| Q \rangle$
\begin{eqnarray}
  \biggl \langle Q \biggl |
  \partial_\nu \left( \hat {\phi}^{\mu\nu} + \hat{\Phi}^{\mu\nu} \right)
  + \epsilon^{3mn} \hat A^m_\nu
  \left(
  \hat F^{n\mu\nu} + \hat G^{n\mu\nu}
  \right)
  \biggl | Q \biggl \rangle & = & 0,
\label{sec4-20a}\\
  \biggl \langle Q \biggl |
  D_\nu \left( \hat F^{m\mu\nu} + \hat G^{m\mu\nu} \right)
  + \epsilon^{3mn}
  \left[
  \hat A^n_\nu \left( \hat {\phi}^{\mu\nu} + \hat \Phi^{\mu\nu} \right) -
  \hat a_\nu \left( \hat F^{n\mu\nu} + \hat G^{n\mu\nu} \right)
  \right]
  \biggl | Q \biggl \rangle & = & 0
\label{sec4-20b}
\end{eqnarray}
One problem in using these equations to obtain expectation
values like $\langle A^m_\mu \rangle$, is that these equations
involve not only powers or derivatives of $\langle A^m_\mu \rangle$
({\it i.e.} terms like $\partial_\alpha \langle A^m_\mu \rangle$ or
$\partial_\alpha \partial_\beta \langle A^m_\mu \rangle$)
but also contain terms like $\mathcal{G}^{mn}_{\mu\nu} =
\langle A^m_\mu A^n_\nu \rangle$. Starting with Eqs. (\ref{sec4-20a})--
(\ref{sec4-20b}) one can generate an operator differential
equation for the product $\hat A^m_\mu \hat A^n_\nu$ thus allowing
the determination of the Green's function $\mathcal{G}^{mn}_{\mu\nu}$.
However this equation will in turn contain other, higher order
Green's functions. Repeating these steps leads to an infinite set
of equations connecting Green's functions of ever increasing
order. This construction, leading to an infinite set of coupled,
differential equations, does not have an exact, analytical solution
and so must be handled using some approximation.
\par
Operators are only well determined if there is a Hilbert space of quantum
states. Thus we need to ask about the definition of the quantum states
$| Q \rangle$ in the above construction. The resolution to this
problem is as follows: There is an one-to-one correspondence
between a given quantum state $| Q \rangle$ and the infinite set
of quantum expectation values over any product of field operators,
$\mathcal{G}^{mn \cdots}_{\mu\nu \cdots}(x_1, x_2 \ldots) =
\langle Q | A^m_\mu (x_1) A^n_\nu (x_2) \ldots
| Q \rangle$. So if all the Green's functions
-- $\mathcal{G}^{mn \cdots}_{\mu\nu \cdots}(x_1, x_2 \ldots)$ --
are known then the quantum states, $| Q \rangle$ are known,
\textit{i.e.} the action of $| Q \rangle$ on any product
of field operators $\hat A^m_\mu (x_1) \hat A^n_\nu (x_2) \ldots$
is known. The Green's functions are determined from the above,
infinite set of equations (following Heisenberg's idea).
\par
Another problem associated with products of field operators
like $\hat A^m_\mu (x) \hat A^n_\nu (x)$ which occur in
Eq. \eqref{sec4-10b} is that the two operators occur at the
same point. For \textit{non-interacting} field it is well
known that such products have a singularity. In this paper
we are considering \textit{interacting} fields so it is
not known if a singularity would arise for such products
of operators evaluated at the same point. Physically
it is hypothesized that there are situations in interacting
field theories where these singularities do not occur
({\it e.g.} for flux tubes in Abelian or non-Abelian theory
quantities like the ``electric'' field inside the tube,
$\langle E^a_z \rangle < \infty$, and energy density
$\varepsilon (x) = \langle (E^a_z)^2 \rangle < \infty$ are
nonsingular). Here we take as an assumption that such singularities
do not occur.
\par
We now enumerate our basic assumptions:
\begin{enumerate}
  \item
  After quantization the components $\hat A^m_\mu (x)$
  become stochastic. In mathematical terms we write this
  assumption as
\begin{equation}
  \left\langle A^m_\mu (x) \right\rangle = 0
  \qquad \text{and} \qquad
  \left\langle
  A^m_\mu (x) A^n_\nu (x) \right\rangle =
  - \varphi (x) \delta ^{mn} \eta _{\mu\nu}
  \label{sec5-10}
\end{equation}
  where $\varphi (x)$ is some scalar field,
  $\eta = \{ +1, -1, -1, -1 \}$. This would give a problem with
  the time components in that $\left\langle A^m_0 A^m_0 \right\rangle < 0$.
  Thus to deal with this we also assume that the fields are static and
  have no time component, {\it i.e.} $A^m_0 = 0$.
  \item
  The components $a^a_\mu$ of the subgroup G can have some order so that
  certain expectation values can have non-zero values, for example
\begin{equation}
  \left\langle H^a_z \right\rangle =
  \left\langle (\nabla \times \vec{a})_z  \right\rangle \neq 0 .
\label{sec5-20}
\end{equation}
   Such conditions are meant to imply that $a_{\mu}$ (or
   certain quantities derived from it) develops a non-zero
   expectation value for some non-trivial, non-vacuum
   boundary conditions ({\it e.g.} the presence of external
   quarks). Such conditions are not connected with vacuum states
   since this would imply a violation of the Lorentz symmetry
   of the QCD vacuum.
  \item
  The gauge potentials $a^a_\mu$ and $A^m_\mu$ are not correlated.
  Mathematically this means that
\begin{equation}
  \left\langle f(a^a_\mu) g(A^m_\nu) \right\rangle =
  \left\langle f(a^a_\mu) \right\rangle
  \left\langle g(A^m_\mu) \right\rangle
\label{sec5-30}
\end{equation}
  where $f,g$ are any functions
\end{enumerate}
These assumptions are a variation of the Abelian Projection
ideas, since there the SU(N)/G components of the gauge fields
are suppressed. The characterization of the off-diagonal fields
as stochastic is a result of the first part of Eq.
(\ref{sec5-10}), $\left\langle A^m_\mu (x) \right\rangle = 0$.
The second part of Eq. (\ref{sec5-10}) is related to some
recent work \cite{gubarev} \cite{kondo2} which demonstrates the
physical importance of the expectation value of the square of the
non-Abelian gauge potential to the dynamics of non-Abelian field
theory. The surprising thing about this is that the non-Abelian
gauge potential (and its square)
is gauge {\it variant}, and one would think that
physical quantities should only be constructed from gauge
{\it invariant} quantities. In previous work \cite{amb}
\cite{cw} \cite{cw1} one had conditions similar to the first
assumption above, but in terms of the expectation values of
the Yang-Mills field strength tensor and its square --
$\langle G_{\mu \nu} \rangle =0$ and
$\langle G_{\mu \nu}^2 \rangle \neq 0$.
One way of looking at the condition,
$\left\langle A^m_\mu (x) A^n_\nu (x) \right\rangle =
- \varphi (x) \delta ^{mn} \eta _{\mu\nu}$, is that it represents
the condensation of the off-diagonal SU(2) gluons into
effective scalar fields, $\varphi (x)$. This provides a physical
motivation for a connection of the present work to
the Ginzburg-Landau model of superconductivity. In
Ginzburg-Landau theory the scalar field represents a
condensation of electrons {\it i.e.} the Cooper pairs.
This association between the expectation of the square
of the off-diagonal gauge potentials with a scalar field
is also similar to ref. \cite{cw} except
there the association was between $\langle G_{\mu \nu}^2 \rangle \neq 0$
and the scalar field.

\section{London's equation}

In this section we want to show how London's equation emerges
from Eqs. (\ref{sec4-10a})-(\ref{sec4-10b}) under the setup
outlined above. London's equation describes the Meissner effect in
ordinary superconductivity. Showing that the same equation
emerges from a quantized non-Abelian gauge theory gives support
to the dual superconducting picture of the QCD vacuum.
Because of the stochastic assumption above we will not be
interested in the off-diagonal components of the gauge fields,
$\left\langle A^m_\mu \right\rangle$. Thus we will not
worry about Eq. (\ref{sec4-10b}) which is the equation that
determines these off-diagonal components. The Abelian field
$a_\mu$ is determined from Eq. (\ref{sec4-10a}) which is
linear in $a_\mu$. Because of this we take the Abelian gauge field
as classical \cite{dzhsin}. This leads to the following equation
\begin{equation}
  \partial_\nu \left(
  \phi^{\mu\nu} + \left\langle \Phi ^{\mu\nu} \right\rangle
  \right) = - \epsilon^{3mn}
  \left(
  \left\langle A^m_\nu F^{n\mu\nu} \right\rangle +
  \left\langle A^m_\nu G^{n\mu\nu} \right\rangle
  \right) .
\label{sec6-10}
\end{equation}
Note the Abelian term, $\phi^{\mu\nu}$, is treated classically while
the remaining terms which involve combinations of the off-diagonal
fields are treated as quantum degrees of freedom via
the expectation values. To calculate these expectation values
we take, as a first approximation, the scalar function of Eq.
(\ref{sec5-10}) as a constant, {\it i.e.} $\varphi (x) = \varphi _0$
\begin{equation}
  \left\langle A^m_\mu (x) A^n_\nu (x) \right\rangle =
  -\varphi_0 \delta^{mn} \eta_{\mu\nu}
\label{sec6-20}
\end{equation}
Then this gives
\begin{eqnarray}
  \left\langle \Phi_{\mu\nu} \right\rangle & = &
  \epsilon^{3mn} \left\langle A^m_\mu A^n_\nu \right\rangle = 0 ,
\label{sec6-30a}\\
  \left\langle A^m_\nu G^{n\mu\nu} \right\rangle & = &
  \epsilon^{np3}
  \left(
  \left\langle A^m_\nu A^{p\mu} \right\rangle a^\nu -
  \left\langle A^m_\nu A^{p\nu} \right\rangle a^\mu
  \right) = -3 \varphi_0 \epsilon^{3mn} a^\mu .
\label{sec6-40}
\end{eqnarray}
The next term is
\begin{equation}
  \left\langle A^m_\nu F^{n\mu\nu} \right\rangle =
  \left\langle A^m_\nu \partial^\mu A^{n\nu} \right\rangle -
  \left\langle A^m_\nu \partial^\nu A^{n\mu} \right\rangle +
  \epsilon^{npq} \left\langle A^m_\nu A^{p\mu} A^{q\nu} \right\rangle .
\label{sec6-50}
\end{equation}
For the disordered, non-diagonal components we will set
$\langle A^{m_1}_{\mu_1}(x) A^{m_2}_{\mu_2}(x) \ldots A^{m_n}_{\mu_n}(x)
\rangle \equiv 0$  if $n$ is odd. For the other
terms in the right-hand-side of Eq. \eqref{sec6-50} we note that
\begin{eqnarray}
  \partial_\alpha \left\langle A^m_\mu A^n_\nu \right\rangle =
  \left\langle \partial_\alpha A^m_\mu A^n_\nu \right\rangle +
  \left\langle A^m_\mu \partial_\alpha A^n_\nu \right\rangle = 0 ,
\label{sec6-60}\\
  \left\langle \partial_\alpha A^m_\mu A^n_\nu \right\rangle =
  - \left\langle A^m_\mu \partial_\alpha A^n_\nu \right\rangle .
\nonumber
\end{eqnarray}
For these stochastic, non-diagonal components the last expression
should not depend on the order of the indices $(m,n)$ and
$(\mu , \nu)$  \textit{i.e.}
$\left\langle \partial_\alpha A^m_\mu A^n_\nu \right\rangle =
\left\langle \partial_\alpha A^n_\nu A^m_\mu \right\rangle =
\left\langle A^m_\mu \partial_\alpha A^n_\nu \right\rangle$.
Using this with Eq. (\ref{sec6-60}) gives
\begin{equation}
\left\langle \partial_\alpha A^m_\mu A^n_\nu \right\rangle =
\left\langle A^m_\mu \partial_\alpha A^n_\nu \right\rangle = 0.
\label{sec6-70}
\end{equation}
Putting all this together gives from Eq. (\ref{sec6-10})
\begin{equation}
  \partial_\nu \phi ^{\mu\nu} = 6 \varphi_0 a^\mu .
\label{sec6-80}
\end{equation}
applying the Lorentz gauge condition, $\partial_\nu a^\nu = 0$,
then yields
\begin{equation}
  \partial_\nu \partial^\nu a^\mu = -6 \varphi_0 a^\mu .
\label{sec6-90}
\end{equation}
This is London's equation for the U(1) ordered phase
in the presence of disordered SU(2)/U(1) phase. To demonstrate
how this leads to a Meissner-like effect for the U(1) gauge
field, $a_{\mu}$, we take half of all space
as being filled by the stochastic phase  ({\it e.g.}
$\varphi (x)= \varphi_0 \neq 0$ for $y>0$
and $\varphi (x) =0$ for $y<0$).
In this case the Abelian gauge field has only a dependence
on $y$, $a_{\mu} (y)$, and Eq. (\ref{sec6-90}) becomes
\begin{equation}
  \frac{d^2 a_\mu}{dy^2} = 6 \varphi_0 a_\mu
\label{sec7-10}
\end{equation}
which has the solution
\begin{equation}
  a_\mu = a_{0 \mu} e^{-\sqrt{6\varphi_0} y} .
\label{sec7-20}
\end{equation}
Thus the magnetic field $H_z = H_{0z} e^{-\sqrt{6\varphi_0} y}$ is
exponentially damped as it penetrates the region with the
stochastic phase.

From eqs. (\ref{sec7-10})-(\ref{sec7-20}) the effective mass
of the Abelian field is $m_{eff} = \sqrt{6 \varphi_0}$. On the
other hand eq. (\ref{sec6-20}) (up to a group factor of
$2/3$) is similar to the relationship given in ref.
\cite{cw1} (see eq. (32) of that reference) between the
effective gluon mass and the expectation of the square
of {\it all} the gauge potentials. Thus from eq.
(\ref{sec6-20}) we find that in our model the effective mass of
the SU(2) gluons associated with the off-diagonal
gauge potentials is different from the effective mass of the
gauge boson associated with the Abelian gauge potential. The
difference in masses between the condensate represented
by the scalar field and the gauge boson is also found in
spontaneous symmetry breaking of a gauge symmetry. For example,
consider the Ginzburg-Landau Lagrangian with an Abelian gauge
field, $A_{\mu}$, and a complex scalar field, $\varphi$,
describing the condensate
\begin{equation}
\label{gl}
{\cal L}_{GL} = (D_{\mu} \varphi)(D_{\mu} \varphi)^{\ast}
- m^2 |\varphi | ^2
-\lambda | \varphi |^4
\end{equation}
where $D_{\mu} =\partial _{\mu} +ie A_{\mu}$. The condensate has
a mass of $m$ while the gauge boson $A_{\mu}$ will acquire a mass
of $\sqrt{\frac{e^2 m^2}{2 \lambda}}$. In our case the condensate
comes from the same set of SU(2) gauge fields as the Abelian
gauge field. The different behavior/roles of the Abelian and
off-diagonal, non-Abelian gauge fields results from using
ideas similar to Abelian Projection through our first assumption
given in eq. (\ref{sec5-10}) above.

\section{Conclusions}

In this paper we have shown how the London equation emerges
from a non-Abelian gauge theory by combining ideas of a
nonperturbative quantization technique pioneered
by Heisenberg and co-workers, with ideas similar to
Abelian Projection. The importance of this is that
the London equation gives a phenomenological description of
the Meissner effect in superconductors, and the vacuum of some
non-Abelian gauge theories ({\it e.g.} QCD) is often modeled as
a dual superconductor in order to explain confinement. In our
approach we split the gauge group (SU(2) in our case) into a 
subgroup (U(1) in our case) and the coset space (SU(2)/U(1) in
our case). The gauge bosons associated with the coset SU(2)/U(1)
were taken to be in an disordered, stochastic phase,
$\left\langle A^m_\mu (x) \right\rangle = 0$. Mathematically
this statement was contained in Eq. (\ref{sec5-10}) where the
scalar field can be compared to the scalar field in the Ginzburg-Landau
treatment of superconductivity. In the Ginzburg-Landau model
the scalar field represents a condensation of electrons into
Cooper pairs. In our work the scalar field can be thought
of as a condensation of gluons. Just as the E\&M field is
excluded from the superconductor, so in our example the diagonal,
Abelian gauge field is excluded from the disordered phase.

There is a difference between Abelian Projection and
the treatment in the present paper. In Abelian Projection the
off-diagonal components are constructed by applying gauge
fixing, but in our case they emerge from applying the
three assumptions given in section III to the dynamical
equations. In the first approximation we have
neglected the dynamical behavior of the stochastic phase,
by setting $\varphi (x) = \varphi _0$. In this way
we obtained an equation for the Abelian components
of the gauge field which was similar to London's equation for
the vector potential in superconductivity theory. Higher
order approximations in the above procedure would result in
higher order powers and derivatives of $\varphi$. This would
hopefully lead to dynamical equations for $\varphi (x)$
similar to the field equations which result from
the Ginzburg-Landau Lagrangian given in eq. (\ref{gl}).
(Note in this regard that there are two scalar fields
in eq. (\ref{gl}), since there $\varphi$ is complex, and
in eq. (\ref{sec5-10}) there are also effectively two scalar
fields coming from $m=n=1$ and $m=n=2$).
In this case one would be able to construct Nielsen-Olesen
flux tube solutions \cite{no}, which would be very suggestive
toward making a firm connection with the dual
superconducting model of QCD. Such a construction of
an effective Ginzburg-Landau equation for
$\varphi$ would be important in bolstering the claim of a
connection between our approach and the dual superconducting
model of the QCD vacuum.

\section{Acknowledgment}

VD is grateful ISTC grant KR-677 for the financial support and
Alexander von Humboldt Foundation for the support of this work.


\begin{thebibliography}{99}

\bibitem{chernodub}
M.N.Chernodub, F.V.Gubarev, M.I.Polikarpov, V.I.Zakharov,
Phys.Atom.Nucl. \textbf{64},  561-573 (2001).

\bibitem{heisenberg}
W. Heisenberg, \textit{Introduction to the unified field theory of
elementary particles.}, Max - Planck - Institut f\"ur Physik und
Astrophysik, Interscience Publishers London, New York, Sydney,
1966; W. Heisenberg, Nachr. Akad. Wiss. G\"ottingen, N8,
111(1953); W. Heisenberg, Zs. Naturforsch., \textbf{9a},
292(1954); W. Heisenberg, F. Kortel und H. M\"utter, Zs.
Naturforsch., \textbf{10a}, 425(1955); W. Heisenberg, Zs. f\"ur
Phys., \textbf{144}, 1(1956); P. Askali and W. Heisenberg, Zs.
Naturforsch., \textbf{12a}, 177(1957); W. Heisenberg, Nucl. Phys.,
\textbf{4}, 532(1957); W. Heisenberg, Rev. Mod. Phys., \textbf{29},
269(1957).

\bibitem{mandl} F. Mandl and G.Shaw,{\it Quantum Field Theory},
Revised Edition (John Wiley \& Sons 1993)

\bibitem{no1} H.B. Nielsen and P. Olesen, Nucl. Phys. \textbf{B144},
376 (1978) 

\bibitem{no2} H.B. Nielsen and P. Olesen, Nucl. Phys. \textbf{B160},
380 (1979)

\bibitem{amb} J. Ambj{\o}rn and P.Olesen, Nucl. Phys. \textbf{B170},
60 (1980)

\bibitem{hooft} G. 't Hooft, Nucl. Phys. \textbf{B190}, 455 (1981).

\bibitem{kondo} Kei-Ichi Kondo, Phys. Rev. \textbf{D57}, 7467 (1998)

\bibitem{gl} V.L. Ginzburg and L.D. Landau, JETP \textbf{20},
1064 (1950)

\bibitem{cw} J.M Cornwall, Phys. Rev. \textbf{D26}, 1453 (1982)

\bibitem{cw1} J.M. Cornwall, Physica A \textbf{158}, 97 (1989)

\bibitem{gubarev} F.V. Gubarev, L. Stodolsky, and V.I. Zakharov,
Phys. Rev. Lett. \textbf{86}, 2220 (2001); F.V. Gubarev and V.I. Zakharov,
Phys. Lett. \textbf{B501}, 28 (2001)

\bibitem{kondo2} Kei-Ichi Kondo, Phys. Lett. \textbf{B514}, 335 (2001)

\bibitem{dzhsin} V. Dzhunushaliev and D. Singleton, Int. J. Theor. Phys.,
\textbf{38}, 2175 (1999); \textit{ibid.} \textbf{38}, 887 (1999).

\bibitem{no} H.B. Nielsen and P. Olesen, Nucl. Phys. \textbf{B61}, 45 (1973)

\end{thebibliography}
\end{document}